\documentclass[10pt,aps,prd,nofootinbib,superscriptaddress,twocolumn,preprintnumbers,balancelastpage]{revtex4}
\usepackage{amssymb,amsmath,latexsym,graphics, graphicx,epsfig,multirow,comment,hyperref,appendix}

\newcommand {\be} {\begin {equation}}
\newcommand {\ee} {\end {equation}}

\newcommand {\bes} {\begin {equation*}}
\newcommand {\ees} {\end {equation*}}
\newcommand {\trit} {^3\text{H}}

\newcommand{\es}[2] {\begin{equation} \label{#1} \begin{split} #2 \end{split} \end{equation}}

\newcommand{\beq}{\begin{equation}}
\newcommand{\eeq}{\end{equation}}

\newcommand {\kms} {\,\,\text{km}/\text{s}}

\newcommand{\V}[1]{{ \bf #1}}

\newcommand{\sun}{\odot}
\newcommand{\earth}{\oplus}
\newcommand{\Min}{\text{min}}

\newcommand{\esc}{\text{esc}}

\newcommand{\CNB}{\text{C}\nu\text{B}}

\begin{document}
\title{Annual Modulation of Cosmic Relic Neutrinos}
\author{Benjamin R. Safdi}
\affiliation{Department of Physics, Princeton University, Princeton, NJ 08544}
\author{Mariangela Lisanti}
\affiliation{Department of Physics, Princeton University, Princeton, NJ 08544}
\author{Joshua Spitz}
\affiliation{Massachusetts Institute of Technology, Cambridge, MA 02139}
\author{Joseph A. Formaggio}
\affiliation{Massachusetts Institute of Technology, Cambridge, MA 02139}

\date{\today}

\begin{abstract}
The cosmic neutrino background ($\CNB$), produced about one second after the Big Bang, permeates the Universe today.
New technological advancements make neutrino capture on beta-decaying nuclei (NCB)
 a clear path forward towards the detection of the $\CNB$.
We show that gravitational focusing by the Sun causes the expected neutrino capture rate to modulate annually.  The amplitude and phase of the modulation depend on the phase-space distribution of the local neutrino background, which is perturbed by structure formation.  These results also apply to searches for sterile neutrinos at NCB experiments.  Gravitational focusing is the only source of modulation for neutrino capture experiments, in contrast to dark-matter direct-detection searches where the Earth's time-dependent velocity relative to the Sun also plays a role.  
\end{abstract}
\maketitle

The cosmic neutrino background ($\CNB$) is a central prediction of standard thermal cosmology~\cite{Dicke:1965zz}.  It is similar to the cosmic microwave background (CMB) as both are relic distributions created shortly after the Big Bang.  However, while the CMB formed when the Universe was roughly 400,000 years old, the $\CNB$ decoupled from the thermal Universe only $\sim$$1$ second after the Big Bang.   

Indirect evidence for the $\CNB$ arises from the contribution of relic neutrinos to the energy density of the Universe.  This affects the abundances of light elements produced during Big Bang nucleosynthesis, anisotropies in the CMB and structure formation (see~\cite{Weinberg:2008zzc} for a review).  However, direct measurements of cosmic neutrinos are made difficult by the low temperature today of the $\CNB$
($T_\nu \approx 1.95$ K), as well as the small interaction cross section and
neutrino masses.      

The relative strength of a $\CNB$ signal depends on the local over-density\footnote{The over-density is defined to be the excess in the local neutrino density relative to the average density in the Universe.} of cosmic neutrinos, which in turn depends on their masses.  The sum of neutrino masses is constrained to be below \mbox{0.66 eV} (95\% C.L) by Planck+WMAP and high-$l$ data, or \mbox{0.23 eV} (95\% C.L) when measurements of baryon acoustic oscillation are included~\cite{Ade:2013zuv}.  Laboratory-based tritium endpoint~\cite{Aseev:2011dq} and neutrinoless double beta-decay experiments~\cite{Auger:2012ar} also have competitive constraints.    Further, the heaviest neutrino mass-eigenstate must be heavier than $\sim$$0.05$ eV to explain neutrino oscillations~\cite{Beringer:1900zz}.  Currently, there are multiple laboratory experiments dedicated to determining the neutrino masses~\cite{deGouvea:2013onf}, including KATRIN~\cite{2013arXiv1307.5486H}, \mbox{Project 8}~\cite{Doe:2013jfe,Monreal:2009za}, and PTOLEMY~\cite{Betts:2013uya}.  

Detecting the $\CNB$ directly would allow us to test a fundamental prediction of thermal cosmology, allowing us to view much further back in time than is possible with the CMB.  
As a result, detecting the $\CNB$ is often referred to as the `holy grail' in neutrino physics.  PTOLEMY is one of the first experiments dedicated to searching for the $\CNB$.  The promise of a relic neutrino experiment on the horizon motivates careful study of the phenomenology of such a signal.

In this Letter, we show that gravitational focusing (GF)~\cite{Lee:2013wza} of the $\CNB$ by the Sun causes the local relic neutrino density to modulate annually.  This modulation, in turn, is expected to give annually modulating detection rates.  As in dark-matter (DM) direct-detection experiments~\cite{Drukier:1986tm,Lee:2013wza}, annual modulation can serve as a strong diagnostic for verifying a potential signal.  
Moreover, an annual modulation measurement could be used to map the local phase-space distribution of relic neutrinos, which is expected to be perturbed by non-linear structure formation~\cite{Ringwald:2004np,Brandbyge:2010ge,VillaescusaNavarro:2012ag,LoVerde:2013lta}.      

The most promising avenue for detecting relic neutrinos is via neutrino capture on beta-decaying nuclei (NCB)~\cite{Weinberg:1962zza}.  In such interactions, a neutrino interacts with a nucleus $N$, resulting in a daughter nucleus $N'$ and an electron:
\es{NCB}{
\nu_e + N \to N' + e^{-} \, . 
}        
The kinetic energy of the electron is $Q_\beta + E_\nu$, where \mbox{$Q_\beta = M_N - M_{N'}$} is the beta-decay endpoint energy and $E_\nu$ is the neutrino's energy.  Note that there is no threshold on $E_\nu$ when the parent nucleus is more massive than the daughter.  As a result, NCB experiments are capable of detecting $\CNB$ neutrinos with 
$E_\nu \lesssim {\cal O}(\text{eV})$. 

The NCB process in~\eqref{NCB} is virtually indistinguishable from the corresponding beta decay.  However, the emitted electron has an energy \mbox{$\geq Q_\beta + m_\nu$}, while the energy is \mbox{$\leq Q_\beta - m_\nu$} for beta decay.  The signal and background are therefore separated by an energy gap of $2 m_\nu$.  For realistic neutrino masses, these experiments must have sub-eV resolution to reconstruct the energy of the final-state electron and discriminate NCB from beta decay.  
 \begin{figure*}[tb]
\leavevmode
\begin{center}$
\begin{array}{lr}
\scalebox{.42}{\includegraphics{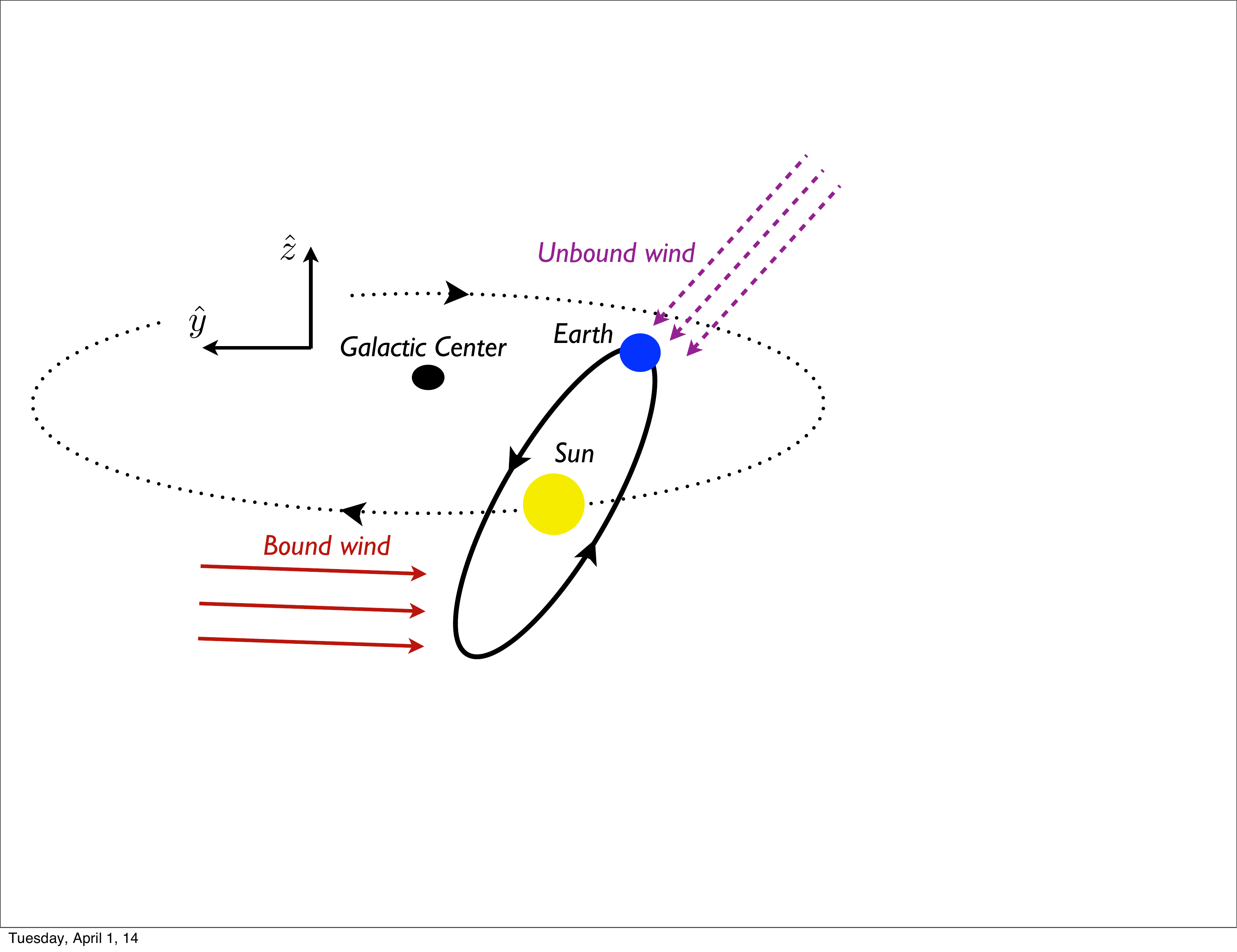}} & \qquad \scalebox{.42}{\qquad \qquad \includegraphics{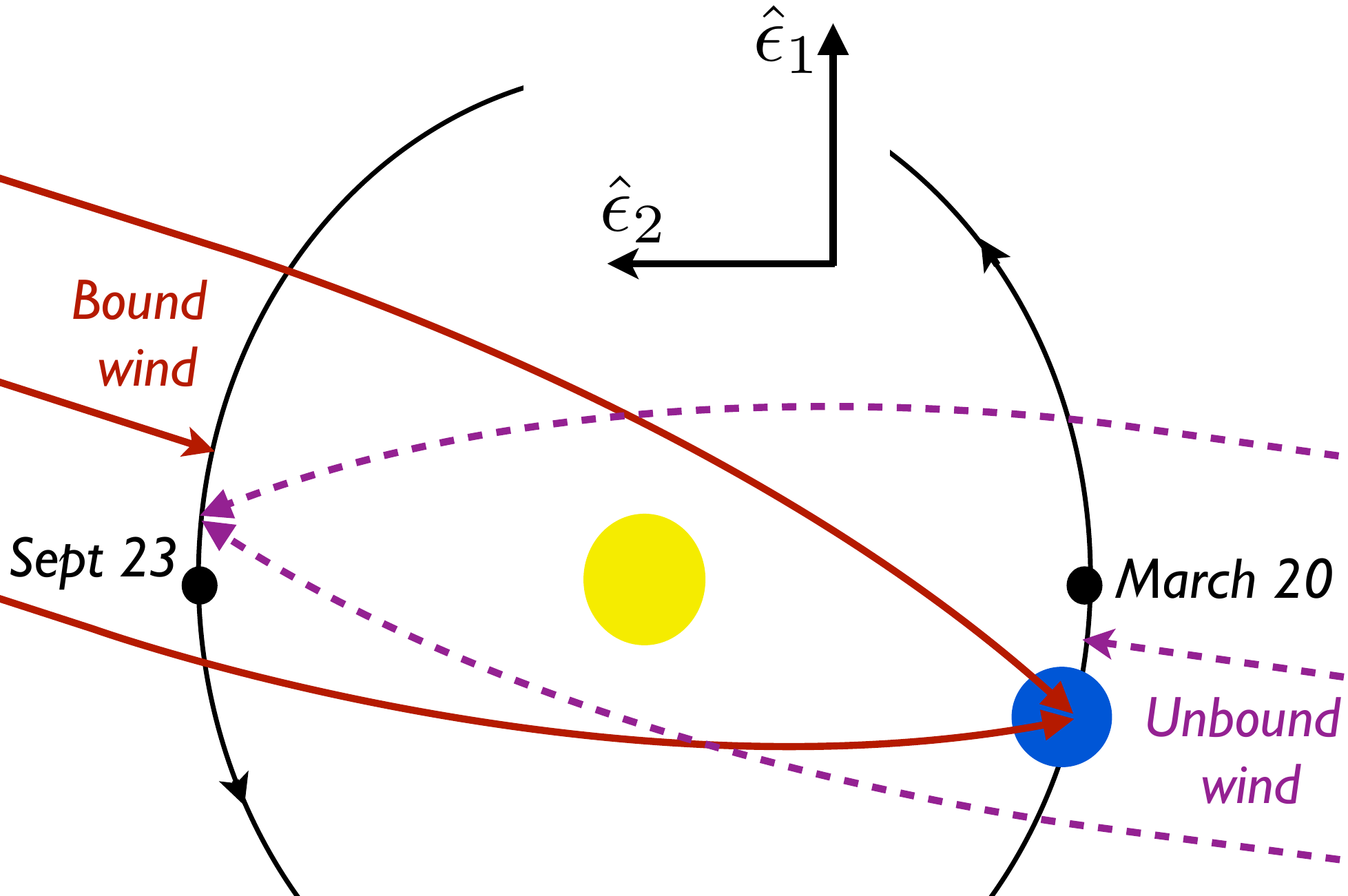}} \\
 \end{array}$
\end{center}
\vspace{-.50cm}
\caption{  
The direction of the neutrino wind relative to the ecliptic plane affects both the amplitude and phase of the modulation.  (left) A projection of the Earth's orbit onto the Galactic ${\bf \hat y}$--${\bf \hat z}$ plane.  The dotted curve illustrates the Sun's orbit about the Galactic Center in the ${\bf \hat x}$--${\bf \hat y}$ plane.  The bound neutrino wind is at an angle $\sim$$60^\circ$ to the ecliptic plane, compared to $\sim$$10^\circ$ for the unbound wind.  This results in a suppressed modulation fraction for the bound neutrinos.    (right) The Earth's orbit in the ecliptic plane, spanned by the vectors ${\bf \hat \epsilon}_1$ and ${\bf \hat \epsilon}_2$.  The focusing of bound and unbound neutrinos by the Sun is also depicted.  The neutrino density is maximal around March~1(September 11) for the bound(unbound) components.  The Earth is shown at March 1 in both panels.
  }
\vspace{-0.15in}
\label{Images}
\end{figure*}   

The neutrino capture rate for an individual nucleus is 
\es{NCBrate}{
\lambda_\nu = \int \sigma_\text{NCB}\, v_\nu \, g_\earth(p_\nu) \, {d^3 p_\nu \over (2 \pi)^3} \,,
}
where $\sigma_\text{NCB}$ is the cross section for~\eqref{NCB}, $v_\nu$ and $p_\nu$ are the neutrino's speed and momentum, respectively, and $ g_\earth(p_\nu)$ is the lab-frame phase-space distribution of neutrinos~\cite{Cocco:2007za}.  The product $\sigma_\text{NCB} v_\nu$ is velocity-independent to very high accuracy when $E_\nu \ll Q_\beta$, which always applies to cosmic neutrinos. 
For tritium decay~\cite{Cocco:2007za}, 
\es{tritiumDecay}{
\sigma_\text{NCB}\left(\trit \right) v_\nu  = ( 7.84 \pm 0.03) \times 10^{-45} \, \, \text{cm}^2 \,.
}
In this limit,~\eqref{NCBrate} simplifies to 
\es{NCBrate_simp}{
\lambda_\nu = {n_\nu } \lim_{p_\nu \to 0} \sigma_\text{NCB} \, v_\nu \,, \quad  n_\nu = \int g_\earth(p_\nu) {d^3 p_\nu \over (2 \pi)^3} \,,
}
where $n_\nu$ is the local neutrino density.

At the time of decoupling, the neutrinos follow the relativistic Fermi-Dirac distribution,
\es{FD}{
\tilde g_{\text{C}\nu\text{B}} (p_\nu) = {1 \over 1+ e^{p_\nu/ T_\nu} } \,,
}  
in the $\CNB$ rest-frame.  Because particle number is conserved after decoupling, this distribution holds even when the neutrinos become non-relativistic, if the effects of cosmological perturbations are  ignored.  In this case, the number density of electron neutrinos today is \mbox{$n_\nu \approx 56$ cm$^{-3}$}.  

While relic neutrinos are relativistic at decoupling, they become non-relativistic at late times and their average velocity is
\es{neutrino velocity}{
\langle v_\nu \rangle = 160 (1 + z) \, (\text{eV} / m_\nu)  \text{ km/s} \,, 
}where $z$ is the redshift and $m_\nu$ is the neutrino mass.    
Galaxies and galactic clusters have velocity dispersions of order $10^2$--$10^3$ km$/$s; dwarf galaxies have dispersions of order 10 km$/$s.  Therefore, sub-eV neutrinos can cluster gravitationally only when $z\lesssim2$.  

In reality, the local neutrino phase-space distribution, as needed for~\eqref{NCBrate},  is more complicated than the Fermi-Dirac distribution.  Non-linear evolution of the $\CNB$ can affect both the density and velocity of the neutrinos today, depending primarily on the neutrino mass~\cite{VillaescusaNavarro:2012ag}.  Ref.~\cite{Ringwald:2004np} simulated neutrino clustering in a Milky Way-like galaxy and found that the local neutrino density is enhanced by a factor of $\sim$$2$($20$) for $0.15$($0.6$) eV neutrinos.  In addition, they find more high-velocity neutrinos than expected from a Fermi-Dirac distribution.

Current numerical predictions for the neutrino phase-space distribution do not account for the relative velocity of the Milky Way with respect to the $\CNB$.  The last scattering surface of cosmic neutrinos is thicker and located closer to us than that for photons, because the neutrinos become non-relativistic at late times~\cite{Dodelson:2009ze}.  
The average distance to the neutrinos' last scattering surface is \mbox{$\sim$2000(500) Mpc} for neutrinos of mass \mbox{0.05(1) eV}~\cite{Dodelson:2009ze}.  For comparison, the last scattering surface for photons is \mbox{$\sim$$10^4$ Mpc} away.  These distances are greater than the sizes of the largest superclusters, which are \mbox{$\mathcal{O}(100)$ \text{Mpc}} in length.  Consequently, it is reasonable to assume that neutrinos do not have a peculiar velocity relative to the CMB.  Measurements of the CMB dipole anisotropy show that the Sun is traveling at a speed of $v_\text{CMB} \approx 369 \kms$ in the direction \mbox{$\V{\hat v_\text{CMB}} = (-0.0695, -0.662 , 0.747)$} relative to the CMB rest-frame~\cite{Kogut:1993ag,Hinshaw:2008kr,Aghanim:2013suk}.  In this Letter, we assume that the same is true for  the $\CNB$ rest-frame.

Given the uncertainties on $g_\earth(p_\nu)$, we consider the limiting cases where the relic neutrinos in the Solar neighborhood are either all unbound or all bound to the Milky Way.  We show that the neutrino capture rate modulates annually in both these limits, but that the modulation phase differs between the two.  More realistically, the local distribution is likely a mix of bound and unbound neutrinos, and the correct phase is different from the examples considered here.

We begin by evaluating the capture rate $\lambda_\nu$ in the limit where all relic neutrinos in the Solar neighborhood are unbound and have not been perturbed gravitationally by the Milky Way or surrounding matter distributions.  In this case, the phase-space distribution at Earth's location is given by~\eqref{FD} in the $\CNB$ rest-frame.  In the non-relativistic limit, the phase-space distribution  can be separated into the density $\rho$ times the normalized velocity distribution $f(\V{v_\nu})$: \mbox{$g(\V{p} = m \V{v_\nu}) = \rho f(\V{v_\nu})$}.
Neglecting gravitational focusing from the Sun, the velocity distribution in the Earth's rest-frame is 
\es{Galboostunbound}{
f_{\earth}(\V{v_\nu}) = \tilde f_{\text{C}\nu\text{B}}(\V{v_\nu} + \V{v_\text{CMB}} + \V{V_\earth}(t)) \, ,  
}
where \mbox{$\V{V_\earth}(t) \approx V_\earth \left( \V{\hat{\epsilon}_1} \cos \omega (t - t_\text{ve}) + \V{\hat{\epsilon}_2} \sin \omega (t - t_\text{ve}) \right)$} is the time-dependent velocity of the Earth with respect to the Sun~\cite{Lee:2013xxa,McCabe:2013kea}.  Note that  $V_\earth \approx 29.79 \kms$, $\omega = 2 \pi/\text{(1 yr)}$, $t_\text{ve}\approx\text{March 20}$ is the time of the vernal equinox, and $\hat{\epsilon}_{1,2}$ are the unit vectors that span the ecliptic plane.  
In this case, the number-density~\eqref{NCBrate_simp} 
is constant throughout the year because the velocity distribution integrates to unity.  As a result, the Earth's time-dependent velocity does not cause the neutrino signal to modulate annually.  

However, the Sun's gravitational field must be accounted for when calculating $n_\nu$.  In the Sun's reference frame, the neutrino distribution appears as a `wind' from the direction $-\V{\hat v_\text{CMB}}$.  The Sun's gravitational field increases the local density when the Earth is downwind of the Sun relative to when it is upwind~\cite{Lee:2013wza}.  The projection of the vector $-\V{\hat v_\text{CMB}}$ to the ecliptic plane determines when the capture rate is extremal.  For unbound neutrinos, the Earth is most upwind of the Sun when
\es{tMin}{
t_\Min \approx t_\text{ve} - {1 \over \omega} \tan^{-1} \left( {\V{\hat v_\text{CMB}} \cdot \V{\hat{\epsilon}_1} \over \V{\hat v_\text{CMB}} \cdot \V{\hat{\epsilon}_2} } \right) \approx t_\text{ve} - 8 \, \, \text{days}\, . 
}
The capture rate is maximal roughly half a year later, around $\sim$September 11.  See Fig.~\ref{Images} for an illustration.
 
Once the Sun's gravitational field is included, the velocity distribution at Earth's location is no longer related to $\tilde f_{\text{$\CNB$}}(\V{v_\nu})$ through a simple Galilean transformation.  Instead,  Liouville's theorem must be used to map the phase-space density at Earth's location to that asymptotically far away from the Sun~\cite{Griest:1987vc,Alenazi:2006wu,Lee:2013xxa}:
\es{frelation}{
\rho \, f_\earth(\V{v_\nu}) = \rho_\infty \, \tilde f_{\text{$\CNB$}} \left(\V{v_\text{CMB}}+ \V{v_\infty} \left[ \V{v_\nu} +  \V{V_\earth(t)} \right]\right) \,.
 }
Note that $\rho_\infty$ is the density far away from the Sun and is different from the local density $\rho$.  In addition, 
 \es{VinftyS}{
 \V{v_\infty}[\V{v_s}] &= {v_\infty^2\V{v_s} + v_\infty(G \, M_\sun / r_s) \V{\hat r_s} - v_\infty \V{v_s} (\V{v_s} \cdot \V{ \hat r_s}) \over v_\infty^2 + (G \, M_\sun / r_s)-v_\infty (\V{v_s} \cdot {\bf \hat r_s}) } \,
 } 
 is the initial Solar-frame velocity for a particle to have a velocity $\V{v_s}$ at Earth's location, where  
$\V{r_s}$ is the position vector that points from the Sun to the Earth~\cite{Lee:2013xxa,McCabe:2013kea}, and conservation of energy gives \mbox{$v_\infty^2 = v_s^2 - 2 \, G \, M_\sun / r_s$}.

\begin{figure}[tb]
\begin{center}
\includegraphics[width=3.5in]{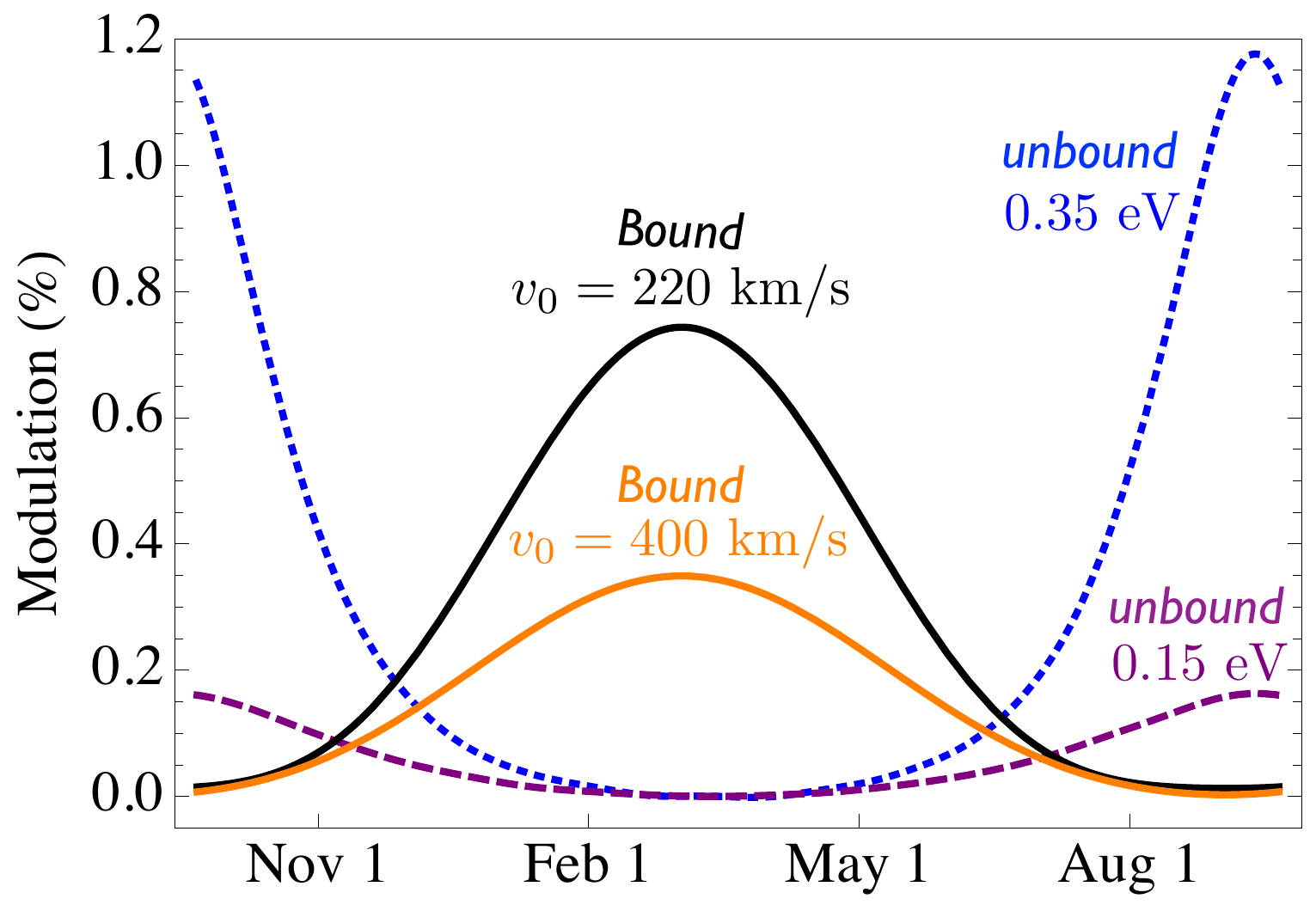}
\end{center}
\vspace{-.50cm}
\caption{The fractional modulation, defined in~\eqref{fracMod}, throughout the year.  The dotted blue and dashed purple curves take the $\CNB$ frame to coincide with the CMB frame and use the Fermi-Dirac distribution~\eqref{FD}.  These calculations neglect the gravitational potential of the Milky Way, which would affect the direction and speed of the unbound wind.  The solid black and orange curves assume that the neutrinos are bound to the Galaxy and use the SHM~\eqref{SHMf}.  More realistically, the phase and amplitude of the modulation will depend on the local fraction of bound versus unbound neutrinos.
  }      
\vspace{-0.15in}  
\label{MWframe}
\end{figure}

The capture rate is obtained by substituting~\eqref{frelation} into~\eqref{NCBrate_simp} and integrating.  The fractional modulation, 
\es{fracMod}{
\text{Modulation} \equiv {\lambda_\nu(t) - \lambda_\nu(t_\Min) \over \lambda_\nu(t) + \lambda_\nu(t_\Min)} \,, }
is shown in Fig.~\ref{MWframe} for  $m_\nu = 0.15$ and $0.35$ eV.  
The maximum modulation fraction for each case is $\sim$0.16\% and $\sim$1.2\%, respectively.  If $m_\nu = 0.6$ eV, the modulation fraction can be  as large as $\sim$3.1\%. 

The effects of GF are most pronounced for slow-moving particles.  These particles spend more time near the Sun and their trajectories are deflected more strongly.  The modulation fraction depends on particle speed as \mbox{$\sim$$\big( v_\esc^S / v_s \big)^2$}, where \mbox{$v_\esc^S \approx 40 \kms$} is the speed to escape the Solar System from Earth's location, and $v_s$ is the particle's Solar-frame speed~\cite{Lee:2013wza}.  When \mbox{$m_\nu = 0.35$ eV}, the mean neutrino speed in the Solar frame is \mbox{$\sim$460 \kms}.  This explains why the modulation fraction is approximately \mbox{$\big(v_\esc^S /  460 \kms \big)^2 \sim0.76\%$}.  On the other hand, when \mbox{$m_\nu = 0.15$} eV, the mean neutrino speed is $\sim$$1100 \kms$ and the modulation fraction is approximately \mbox{$\big( v_\esc^S /  1100 \kms \big)^2 \sim 0.13\%$}.

Next, we consider the case of relic neutrinos bound to the Milky Way.  We assume that these neutrinos have sufficient time to virialize and that their Galactic-frame velocity distribution $\tilde f(\V{v_\nu})$ is isotropic.  Regardless of the exact form of $\tilde f(\V{v_\nu})$, the clustered-neutrino `wind' in the Solar frame is in the direction $-\V{\hat v_\sun}$, where \mbox{$\V{v_\sun} \approx (11 ,232 ,7 ) \kms$} is the velocity of the Sun in the Galactic frame~\cite{Schoenrich:2009bx}.  The capture rate is minimal at
\es{tMin}{
t_\Min \approx t_\text{ae} - {1 \over \omega} \tan^{-1} \left( {\V{\hat v_\odot} \cdot \V{\hat{\epsilon}_1} \over \V{\hat v_\odot} \cdot \V{\hat{\epsilon}_2} } \right) \approx t_\text{ae} - 19 \, \, \text{days}\, ,
}
where $t_\text{ae}$ is the autumnal equinox.  The date of maximal rate is half a year later $\sim$March 1, as shown in Fig.~\ref{Images}.

The velocity distribution $\tilde f(\V{v_\nu})$ determines the shape and the amplitude of the modulation.  For a given velocity distribution, the fractional modulation~\eqref{fracMod} is computed using~\eqref{frelation}, with the obvious substitutions.  As an example, we let the clustered-neutrino velocity distribution at the Sun's location follow that of the DM halo.  The DM velocity distribution is typically modeled by the Standard Halo Model (SHM)~\cite{Drukier:1986tm}, an isotropic Gaussian distribution with a cut-off at the escape velocity \mbox{$v_\esc \approx 550 \kms$}~\cite{Smith:2006ym}:
 \es{SHMf}{
\tilde f (\V{v_\nu}) = \left\{ \begin{array}{ll}
{1 \over N_\esc } \left( {1 \over \pi v_0^2 } \right)^{3/2} e^{- \V{v_\nu}^2 / v_0^2 } \, \, \,  \, \, \,&|\V{v_\nu}| < v_\esc \\
0 \, \qquad &|\V{v_\nu}| \geq v_\esc \,,
\end{array}
\right.
}
 with $N_\esc$ a normalization factor.  For DM, the dispersion \mbox{$v_0 = 220 \kms$} is usually taken to be the speed of the local standard of rest relative to the Galactic Center.  However, we also consider the case when \mbox{$v_0 = 400 \kms$} because numerical simulations of neutrino clustering suggest that bound neutrinos may have faster speeds than their DM counterparts~\cite{Ringwald:2004np,Brandbyge:2010ge,VillaescusaNavarro:2012ag,LoVerde:2013lta}.  Fig.~\ref{MWframe} shows the fractional modulation for the clustered neutrinos.  The maximum modulation fraction is $\sim$0.75\% and $\sim$0.35\% for $v_0 = 220$ and $400 \kms$, respectively. 
 
The amplitude and phase of the modulation depend on the neutrino's mass, as well as the fraction of cosmic neutrinos that are bound versus unbound to the Galaxy.  For the examples we have discussed, the modulation can be a \mbox{$\sim$$0.1$--$1$\%} effect.  How much exposure would a tritium-based NCB experiment need to detect this modulation?  For a neutrino number density of $\bar n_\nu \approx 56$ cm$^{-3}$ and the capture cross section given in~\eqref{tritiumDecay}, such an experiment should observe $\sim$$100 $ events per kg-year. If annual modulation is a $0.1$--$1$\% effect, $\sim$$10^4$--$10^6$ events are needed to detect it with roughly two-sigma significance, in consideration of statistical uncertainties only.

This estimate depends however on the over-density of clustered neutrinos, which is not well-understood.  Numerical simulations currently find that $\mathcal{O}$(10) over-densities are feasible, but further study is needed.  An accurate prediction of the neutrino phase-space distribution at Earth's location (and thus the phase and amplitude of the modulation) requires a dedicated simulation that properly accounts for the motion of the Milky Way with respect to the C$\nu$B, the embedding of the Milky Way in the local supercluster, and the location of the Sun within the Milky Way.    

To assess the experimental implications of a modulating signal, consider the PTOLEMY experiment, which plans to use a surface-deposition tritium target with total tritium mass of \mbox{$\sim$100 g}~\cite{Betts:2013uya}.  This is a significant increase in scale from the KATRIN experiment, which has a gaseous tritium target with an effective mass of \mbox{66.5 $\mu$g}~\cite{Kaboth:2010kf}.  Assuming no clustering, PTOLEMY should observe $\sim$$10$ events per year due to $\CNB$ neutrinos.  This will provide the first detection of the unmodulated cosmic neutrino rate, but will not suffice to detect an annual modulation.  In other words, this experiment will measure the neutrino over-density, but it will not be able to probe the velocity distribution.

If the local neutrino density is enhanced by a factor of $\sim$$10^3$ or more, then PTOLEMY may be able to detect annual modulation within a year.  Such large over-densities can arise, for instance, in models where neutrinos interact via a light scalar boson, forming neutrino ``clouds"~\cite{Stephenson:1996qj}.  Because PTOLEMY uses atomically-bound tritium, it is feasible to scale up to
a $\sim$$10$ kg-sized target or larger consisting of multiple layers of graphene substrate~\cite{tully}. The next-generation experiments may be sensitive to a modulating neutrino signal, even if the local neutrino over-density is negligible.
  
Tritium-based NCB experiments will also be sensitive to relic sterile neutrinos.  A number of anomalies in ground-based neutrino experiments point towards a sterile neutrino with ${\cal O}(\text{eV$^2$})$ mass-squared splitting from the active-neutrino eigenstates and sterile-electron-neutrino mixing parameter \mbox{$|U_{e4}|^2 \sim 10^{-3}$--$10^{-1}$}~\cite{Aguilar:2001ty,Mention:2011rk,Aguilar-Arevalo:2013pmq}.  Moreover, if the recent B-mode power-spectrum measurements by BICEP2~\cite{Ade:2014xna} are interpreted as being produced by metric fluctuations during inflation, then an analysis of the combined Planck+WMAP+BICEP2 data prefers $N_\text{eff} = 4.00 \pm 0.41$ (68\% C.L.)~\cite{Giusarma:2014zza}, suggesting the presence of an extra light species.\footnote{Several other global analyses that include the BICEP2 data have also been published recently~\cite{Zhang:2014dxk,Dvorkin:2014lea}.  All find that the CMB and BICEP2 results can be made consistent with the addition of a light, sterile neutrino.}

The morphology of a relic sterile neutrino signal at an NCB experiment is similar to that of the active neutrinos.  The detection rate is suppressed by $|U_{e4}|^2$ because the
mostly-sterile fourth mass eigenstate contributes to the electron
energy spectrum through its electron-flavor component.  However, the local over-density of the fourth mass eigenstate may be greater than that of the active neutrinos if the new state is significantly more massive.  PTOLEMY will be sensitive to a portion of the sterile-neutrino parameter space suggested by the ground-based neutrino experiment anomalies; a $\sim$$10$ kg-sized target would cover the entire parameter space.  
  
Scenarios where the DM is a sterile neutrino~\cite{Dodelson:1993je} (for reviews, see~\cite{Boyarsky:2009ix,Kusenko:2009up,Boyarsky:2012rt}) with small mixing to the electron neutrino may also be probed at NCB experiments~\cite{Li:2010vy}.  
A sterile neutrino DM signal should also modulate annually. 
 If the local velocity distribution is modeled by the SHM with $v_0=220\text{ km/s}$, the modulation amplitude is the same as the corresponding line for bound relic neutrinos in Fig.~\ref{MWframe}.  

Assuming the sterile neutrinos constitute all of the DM density (\mbox{$3$ GeV\,cm$^{-3}$}), a tritium-based detector should observe 
\es{observedS}{
  \sim4 \times 10^5 \,\,|U_{e4}|^2 \,\left({ \text{keV} \over  m_{\nu_4}} \right) \, \, { \text{events} \over \text{kg-year}} 
  }
interactions with the DM halo~\cite{Li:2010vy}, where $m_{\nu_4}$ is the sterile-neutrino mass.  This estimate assumes that the capture cross section is given by~\eqref{tritiumDecay} and is independent of velocity.  However, $\sigma_\text{NCB} v_\nu$ depends on $E_{\nu}$ at energies above $\sim$$Q_\beta$~\cite{Cocco:2007za}.  For example, $\sigma_\text{NCB} v_\nu$ is enhanced by an additional factor of $\sim$2(10) if \mbox{$E_{\nu} = 50(500)$ keV}.
  
Strong constraints on sterile neutrino DM  arise from a variety of astrophysical observations, such as the Tremaine-Gunn bound, the Lyman-$\alpha$ forest, and x-ray observations (see ~\cite{Kusenko:2009up,Boyarsky:2012rt} and references therein).  To be consistent with observations, \mbox{$m_{\nu_4} \gtrsim 1$ keV} and \mbox{$|U_{e4}|^2 \lesssim 10^{-5}$}~\cite{Tremaine:1979we, Li:2010vy}.  
Recently, there have been anomalies in the observed x-ray spectrum consistent with sterile neutrino DM of mass $m_{\nu_4} \approx 7$ keV~\cite{Bulbul:2014sua,Boyarsky:2014jta}. 

Detecting sterile neutrino DM with NCB requires an extremely large tritium target mass.  However, it may be easier to scale up a dedicated  ${\cal O}(\text{keV})$ sterile-neutrino DM experiment because the energy resolution of the detector can be relaxed compared to that needed for the detection of relic neutrinos. One promising avenue is to use multiple overlapping layers of titanium-held tritium~\cite{tully}.  Moreover, if such a machine was designed to probe energies $\sim$300 keV, it would also function as a real-time solar $pp$ neutrino detector~\cite{tully}. 
 
In conclusion, we have shown that a cosmic neutrino signal in tritium-based NCB experiments should modulate annually.  The phase and amplitude of the modulation varies depending on the neutrino's mass, which affects how strongly its distribution is perturbed during structure formation.  For the examples that we considered, the modulation fraction can be $\sim$0.1--1\%.  
Annual modulation will first be useful as a method for distinguishing a potential signal from background.  Beyond this stage, annual modulation will be a powerful tool for studying the underlying neutrino velocity distribution.  

It is interesting to contrast the modulation of relic neutrinos with that of weakly interacting DM at direct-detection experiments.  In the latter case, annual modulation is primarily due to the changing velocity of the Earth throughout the year~\cite{Drukier:1986tm}, which causes the rate to be extremized around June~1.  An additional modulation effect due to GF is also present~\cite{Lee:2013wza}, resulting in a shift in the expected phase.  
The DM scenario is in stark contrast to the modulation of $\CNB$ neutrinos discussed in this Letter.  For NCB experiments, GF is the dominant source of annual modulation because $\sigma_\text{NCB} v_\nu$ is nearly velocity-independent over the relevant momentum range.       

Over the years, the promise of a modulating dark-matter signal has guided both experimental design and data interpretation.  The same can be true in the search for relic neutrinos and the development of next-generation $\CNB$ experiments.  If detecting the $\CNB$ is the `holy-grail' of neutrino physics, then $\CNB$ annual modulation is the `Excalibur.'\\

 \noindent
{\it We are especially grateful to 
Christopher Tully for discussing the PTOLEMY experiment with us.  We also thank Matthew Buckley, Francis Froborg, Samuel Lee, David McGady, and Matthias Zaldarriaga for useful discussions.  BRS is supported by the NSF grant PHY-1314198.  JS is supported by a Pappalardo Fellowship in Physics at MIT and by
the NSF grant PHY-1205175. JAF is supported in part by the NSF grant PHY-1205100.} 

\vspace{0in}
\onecolumngrid
\vspace{0.3in}
\twocolumngrid
\def\bibsection{} 
\bibliographystyle{apsrev}
\bibliography{cosmicNeutrinos}

\end{document}